\documentstyle[editedvolume,psfig]{crckapb} 

\begin{opening}
\title{Ram-Pressure Stripping on Dwarf Galaxies}

\author{M.A. G\'omez-Flechoso}
\institute{Observatoire de Gen\`eve\\
           CH-1290 Sauverny (Switzerland)}

\end{opening}

\begin{document}
\begin{abstract}
We use SPH/N-body simulations to study the ram-pressure stripping of the
gas of dwarf galaxies orbiting in a disc galaxy.
The effects of the gaseous 
disc and halo on the satellites are analysed and the results are compared
with the observations of the dwarf spheroidals
(and dwarf irregulars) in the Local Group. The 
stripped gas is compared with the High Velocity Clouds. Finally, a model
of evolution of the dwarfs is proposed.
\end{abstract}

\section{Introduction}
 Some satellite dwarf galaxies in the Local Group (LG) have
low or null gas content (Mateo 1998). However, nearly half of them
are (dynamically) associated with large reservoirs of atomic
gas, not always placed at the dwarf center
(e.g., Carignan 1999).
A correlation between the dwarf gas 
content and its distance to the center of the 
main galaxy (in the LG) exists (Blitz \& Robishaw 2000).
The gas reservoirs of the dwarfs and the High Velocity 
Clouds (HVCs) also seem to be related, as
they both have similar properties 
(Blitz \& Robishaw 2000; Braun \& Burton 1999, 2000).

 In this paper, we study the evolution of a dwarf
irregulars (dIrrs) into dwarf spheroidal (dSph),
when the ram-pressure removes the gas of the dIrrs as
they orbit in a disc galaxy.
This gas can be identify as a HVC. We also present a 
semi-analytical method to simulate the stripping process.

\section{The ram-pressure stripping}
The gas of a satellite moving through a gaseous environment (halo) 
feels a pressure $P\propto \rho_e v^2$, ($\rho_e$ is the halo gas density
and $v$ the satellite velocity) that can remove the gas
from the satellite (Gunn \& Gott 1972). For a satellite cross section $S$ and
a satellite gas mass $M_g$, the ram-pressure stripping of the satellite gas 
is 
\begin{equation}
\vec{a}_{rp}=\frac{-C_D \rho_e S v \vec{v}}{2 M_{g}}
\label{ramacc}
\end{equation}
where $C_D\sim 0.2-1.0$ is a drag parameter. 

Taking into account this acceleration and the satellite 
gravity, the minimum gas density of the halo for the 
total satellite gas stripping is
$\rho_e > \alpha G \Sigma_s \Sigma_g/v^2$,
where the parameter $\alpha \sim 1$ depends on the
geometry of the satellite and $\Sigma_s$ 
($\Sigma_g$) is the satellite surface densities of the stars (gas)
(Takeda et al 1984; Mori \& Burkert 2000; Blitz \& Robishaw 2000).
To obtain a detailed description of the stripping,
hydrodynamical simulations of the satellite+halo system are needed.
Such simulations are time expensive because they 
demand a large particle number to describe all the interactions between 
the satellite and halo and, so far, only the case of a satellite with 
uniform velocity moving through an uniform halo has been considered
(Abadi et al 1999; Mori \& Burkert 2000; Quilis et al 2000). 

\section{Model of the ram-pressure stripping}
In order to reduce the calculation time of the simulations, we have 
described analytically, using a continuous function, the effect of the 
environment on the (N-body) satellite.
We have run simulations using a treesph code (see Barnes \& Hut 1986 and Fux 
1997a for details), modified to include
the external forces.
The gravitational force of the environment affects all the satellite particles
and is obtained from the analytical expression of the total environmental density.
The ram-pressure acceleration only acts on the gas particles $i$, as
  \begin{figure}
  \centerline{
  \psfig{file=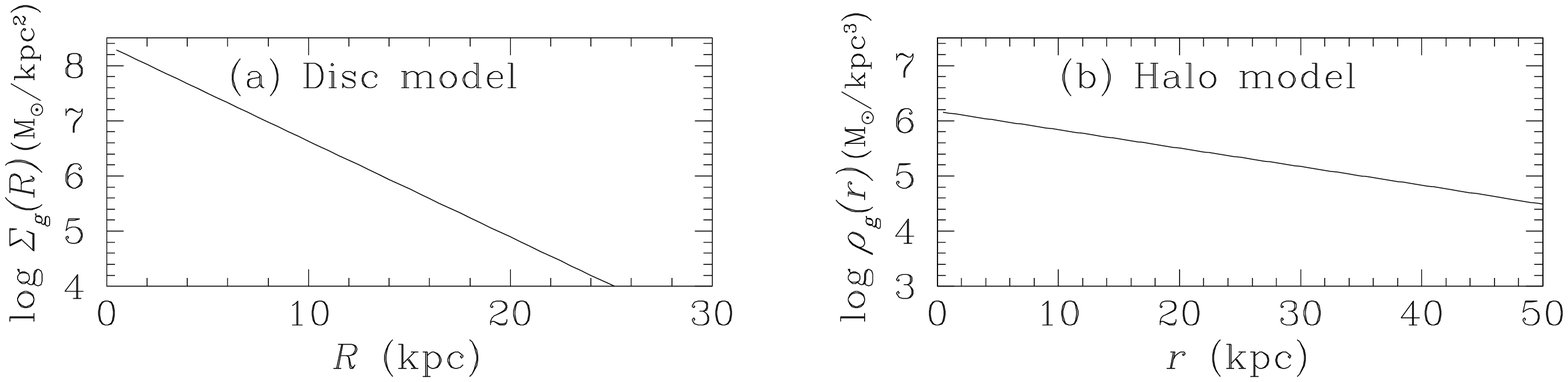,width=\textwidth,angle=0,clip=}%
  }
  \caption{(a) Gas surface density of the disc model, $\Sigma_g$, as function of 
  the cylindrical radius, $R$, and (b) gas density of the halo model, $\rho_g$, 
  as function of the spherical radius,~$r$}
  \label{fig1}
  \end{figure}
\begin{equation}
\vec{a}^{env}_{rp,i} = \frac{-C_D \rho_e \pi h^2_i v_i \vec{v}_i}{2 m_i N^{1/3}_i}
\label{ramacci}
\end{equation}
obtained from eq. (\ref{ramacc}), where $h_i$ is the SPH softening length, $N_i$ the number of gas neighbours,
$m_i$ the mass and $\vec{v}_i$ the velocity of the $i$ particle. 
In the satellite low density regions, $h_i$ increases and 
$N_i$ decreases, therefore the ram-pressure grows 
as expected. The contrary occurs in the satellite high 
density regions. 
Details of the model 
are given in G\'omez-Flechoso (2000).

\section{Simulations of the satellite accretion}
We have run simulations of an analytical multicomponent (halo+bulge+disc) 
galaxy which accretes a dwarf satellite. 
The main galaxy represents a Milky Way like galaxy (see Fux 1997b for details). 
The gas of the primary galaxy is placed either in the disc or in the halo 
(densities are plotted in Fig.~\ref{fig1}).
Our satellites are N-body realizations of a
King model (dimensionless central potential $W_o=4$ and core radius $r_o=1$ kpc),
but we include 10\% of its mass as gas in hydrostatic equilibrium. 
We have considered two
satellites: (i) a Large Magellanic Cloud (LMC) like model
(model A), massive ($M_{sat}=2\times 10^{10}$ M$_{\odot}$) and
dense, and (ii) a Sagittarius like model (model B), with low density 
and $M_{sat}=2\times 10^8$ M$_{\odot}$.

In order to give limits to the gas stripping, we have taken a very
eccentric orbit perpendicular to the disc 
(apocenter $R_a=70$ kpc and pericenter $R_p=8$ kpc), as the stripping 
is higher in high density regions.

\section{Results}
We have study the tidal stripping by a disc galaxy and the ram-pressure 
stripping by either a disc or a halo of gas on a dwarf satellite model.
  \begin{figure}
  \centerline{
  \psfig{file=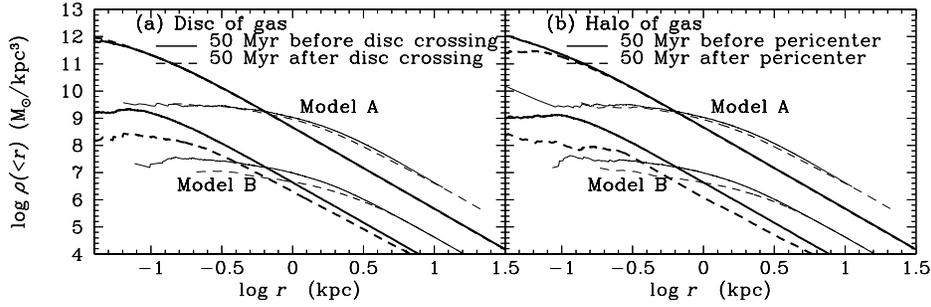,width=\textwidth,angle=0,clip=}%
  }
  \caption{(a) Gas (thick line) and star (thin line) density profiles 
  for the satellite model A (four upper profiles) and model B
  (four lower profiles), 50 Myrs before (solid line) and
  after (dashed line) disc crossing, when the gas of the primary galaxy is
  placed in the disc. (b) Idem that (a) but the gas of the primary galaxy is
  in the halo.}
  \label{fig2}
  \end{figure}

(a) {\it Ram-pressure stripping by a disc:}
The dwarf gas stripping at the disc crossing depends
on the dwarf density. Whereas the satellite model A suffers
very little stripping, model B 
loses $\sim 50 \%$ of the gas mass after the 
first disc crossing and the stripped gas mainly finishes 
bound to the disc, orbiting with the disc circular velocity. 
However, the tidal stripping of the stars
is less significative (density profiles are plotted in Fig. \ref{fig2}a).

(b) {\it Ram-pressure stripping by a halo:}
The ram-pressure stripping occurs at every point along 
the orbit, but mainly at the pericenter (where the halo density is higher).
As in the previous case, the gas of the model A only suffers a slow evolution.
On the contrary, the model B loses
almost all the gas ($\sim 70\%$) soon after the first
pericenter, that forms tails and condensations, whose central
surface density and the velocity ($\sim 10^{20}$ 
atoms/cm$^2$ and $100-300$ km/s, respectively) are similar to those
of the HVCs.
The tidal stripping is not
strong enough to disrupt the satellite, given rise to a dwarf galaxy without
gas (density profiles are plotted in Fig. \ref{fig2}b).

\section{Conclusions}
The evolution of a gaseous dwarf galaxy depends on its density:

(a) High density satellites suffer slow evolution
and marginal gas losses. That can explain the existence
of satellites (as the LMC) with gaseous tails.

(b) Low density satellites lose their gas
after a few pericenters, whereas a large percentage of the stellar
(and dark) material remains bound to the dwarf, supporting
the evolution from dIrrs into dSphs
when they orbit in a disc galaxy. It can also explain the apparent 
correlation between the dwarf gas content and its distance to the primary.

The fate of the stripped gas depends on the distribution of 
the gas in the primary:

(a) If the gas of the primary is in the halo, the stripped gas of the
dwarf forms condensations that can be identified as HVCs.

(b) If the gas is in the disc, the satellite gas
is stripped as it crosses the disc and it finishes orbiting in it.
It can feed the star formation in the disc.

In a general case, both process (stripping by a halo and by a disc) 
coexist, increasing the gas loss rate of the dwarf.

\end{document}